\documentclass{sig-alternate}

\usepackage{algorithm}
\usepackage{algpseudocode}
\usepackage{nccbbb}
\usepackage{tabu}
\usepackage{balance}

\begin{document}

\setcopyright{acmcopyright}


%

\CopyrightYear{2016}
\setcopyright{acmcopyright}
\conferenceinfo{MM '16,}{October 15-19, 2016, Amsterdam, Netherlands}
\isbn{978-1-4503-3603-1/16/10}\acmPrice{\$15.00}
\doi{http://dx.doi.org/10.1145/2964284.2964335}
\clubpenalty=10000
\widowpenalty = 10000

\title{Time Matters: Multi-scale Temporalization of \\Social Media Popularity}
%
%
\author{Bo Wu$^{1,2}$, Wen-Huang Cheng$^{3}$, Yongdong Zhang$^{1}$, Tao Mei$^{4}$\\
$^{1}$Key Laboratory of Intelligent Information Processing, \\
Institute of Computing Technology, Chinese Academy of Sciences, China\\
$^{2}$University of Chinese Academy of Sciences, China\\
$^{3}$Research Center for Information Technology Innovation, Academia Sinica, Taiwan\\
$^{4}$Microsoft Research, China\\
\{wubo, zhyd\}@ict.ac.cn; whcheng@citi.sinica.edu.tw; tmei@microsoft.com\\
}

\maketitle

\begin{figure}[t]
  \centering\includegraphics[width=2.9in]{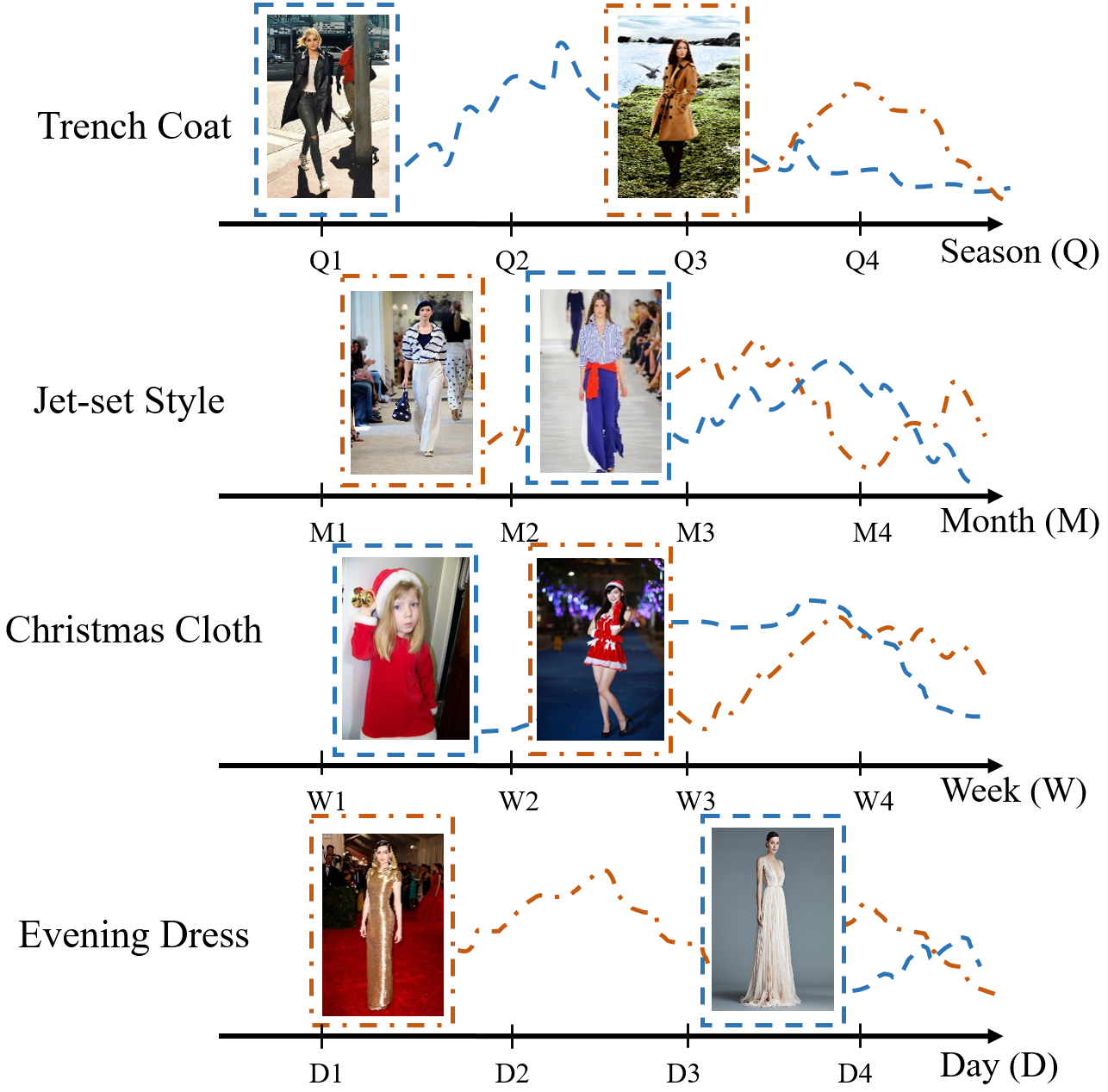}
  \vskip -10pt
   \caption{Social media popularity changes over time and often exhibits varying temporality across different time scales. Clothing fashion, as a typical example, gives a vivid expression of the temporal characteristics for social media popularity. The life spans of four particular fashions are observed to be diversified from few days (e.g., evening dress), specific weeks (e.g., Christmas cloth), several months (e.g., Jet-set style), to certain seasons (e.g., trench coat). Each dashed curved line after an image represents the life cycle of popularity of the corresponding clothes.}\label{challenges}
  \vskip -10pt
\end{figure}

\begin{abstract}
The evolution of social media popularity exhibits rich temporality, i.e., popularities change over time at various levels of temporal granularity. This is influenced by temporal variations of public attentions or user activities. For example, popularity patterns of street snap on Flickr are observed to depict distinctive fashion styles at specific time scales, such as season-based periodic fluctuations for Trench Coat or one-off peak in days for Evening Dress. However, this fact is often overlooked by existing research of popularity modeling. We present the first study to incorporate multiple time-scale dynamics into predicting online popularity. We propose a novel computational framework in the paper, named Multi-scale Temporalization, for estimating popularity based on multi-scale decomposition and structural reconstruction in a tensor space of user, post, and time by joint low-rank constraints. By considering the noise caused by context inconsistency, we design a data rearrangement step based on context aggregation as preprocessing to enhance contextual relevance of neighboring data in the tensor space. As a result, our approach can leverage multiple levels of temporal characteristics and reduce the noise of data decomposition to improve modeling effectiveness. We evaluate our approach on two large-scale Flickr image datasets with over 1.8 million photos in total, for the task of popularity prediction. The results show that our approach significantly outperforms state-of-the-art popularity prediction techniques, with a relative improvement of $10.9\%$--$47.5\%$ in terms of prediction accuracy.
\end{abstract}

%
%
%
%

%
%

%
%

%
%


\keywords{Social media popularity; multi-scale temporal modeling; tensor decomposition and reconstruction; popularity prediction}

\vskip 16pt

\section{Introduction}

Social media has changed the way human interact and has connected billions of users worldwide with friends and family through online content posting. Meanwhile, a plethora of information is generated for competing their attention, e.g., 1.3 million pieces of content are shared on Facebook every single minute~\cite{Bagadiya2016}. Only few posts have the opportunity to become popular while most are ignored over time~\cite{Khosla2014, Wu2016}. Thus, modeling the popularity evolution and predicting the individual popularity (the popularity of a specific post or item) are very active fields of social media research and can lead to a number of applications, such as content recommendation~\cite{McParlane2014, Chi2015}, advertisement placement~\cite{Li2015}, information retrieval~\cite{Wu2014, Tsai2011} and online media caching services~\cite{Wang2013}.

The existing studies of exploiting temporal information for popularity modeling can be grouped into two main categories. The first category builds popularity models to discover general patterns for temporal trends in popularity data~\cite{Kong2014, Roy2013, Szabo2008, Mathioudakis2010, Yang2011, Zhuang2011}. The statistics are typically computed over the entire popularity data, such that specific characteristic needed for individual popularity prediction is often lost. The second category is to estimate the individual popularity of online content by temporal factors or variables with intrinsic time information~\cite{Wu2016, Li2015, Shen2014, He2014, Zhao2015}. Characteristics of time-dependent factors are explicitly incorporated into prediction, and most of the recent researches in this field fall into this category.

Popularities are mirror of society, reflecting shifts in people's preferences in the context of social alteration. Taking clothing fashion as an illustration, images or videos about trendy dresses naturally receive much more attention than out-of-date clothes on social media~\cite{Hidayati2014, Hidayati2012}. Also, a fashion can have a short or long life cycle. For example, as shown in Figure 1, certain types of clothing fashion tend to last for a period that is as short as only few days to weeks, while others will decline slowly or even turn into what is known as a timeless classics. These observations reveal an essential characteristic of popularities, i.e., popularities not only change over time but also operate across a range of different time scales. Recently, several researchers also found similar phenomena on different social platforms influenced by various temporal factors~\cite{Preum2015, Wu2016}, such as user habits, timely events, headline news, and popular products. The underlying momentum is related to temporal variations in the life spams of human activities and preferences. However, to the best of our knowledge, little work has been done to leverage multi-scale temporal dynamics for popularity prediction.

In this paper, therefore, we propose to understand popularity dynamics from a multiple time-scale perspective, named \emph{Multi-scale Temporalization} (MT), which is a structured modeling for popularity dynamics into multiple levels of time scale. As shown in Figure 2, it is distinctive from the conventional paradigm that considers all the popularity data as a whole popularity tensor~\cite{Wu2016, Cappallo2015, Kong2011}. We attempt to explore them by considering multi-scale effects together over time, simultaneously in multiple time scales. As preprocessing, we design a data rearrangement step to enhance contextual relevance of neighboring data for reducing the noise caused by context inconsistency. Then we propose a multi-scale tensor decomposition with considering temporal information and context information in three-dimensional data space of popularity (formed by the user, post, and time dimensions). Finally, we perform reconstruction through minimizing a joint low-rank constraint for predicting online popularity. Therefore, our approach can leverage multiple levels of temporal characteristics and reduce data noise to improve the prediction effectiveness.

In the experiments, we demonstrate the effectiveness of the proposed framework through experiments with two large-scale Flickr datasets for the task of popularity prediction. One of the datasets is from VSO~\cite{Borth2013} and the other is crawled from Flickr, with over 1.8 million photos in total. The results show that our approach significantly outperforms state-of-the-art popularity prediction techniques, with a relative improvement of averagely 10.9\%--47.5\% in prediction accuracy. Moreover, the proposed framework is novel and can be readily extended to other content domains and application scenarios with the analysis of various evolutionary dynamics.

The main contributions of this study are: (i) we address popularity prediction within the multiple time-scale perspective; (ii) we propose a framework MT to predict online popularity based on decomposition via joint low-rank constraints; (iii) we also propose a data rearrangement strategy based on the aggregation of context information for reducing the noise caused by context inconsistency; (iv) we evaluate our approach on large scale datasets and achieve a significant outperformance over the state-of-the-art methods.

The rest of the paper is organized as follows. Related approaches of temporal modeling for popularity are discussed in Section 2. We propose the multi-scale temporalization model in Section 3. In Section 4, we report experimental evaluations of the proposed approach and comparisons with state-of-the-art algorithms. Finally, the conclusions and future work are presented in Section 5.

\section{Related Work}
The temporal information modeling for popularity dynamics on social media has received much research attention recently. Existing works can be concluded into two main paradigms.

The first paradigm focuses on exploring the patterns of evolving trend for popularity variations ~\cite{Kong2014, Roy2013, Yang2011, Mathioudakis2010}. Throughout the early years, Szabo and Huberman attempted to analyze online popularity trend growth with pattern characteristics~\cite{Szabo2008}. Mathioudakis and Koudas detected popularity trends by bursty key words in Twitter~\cite{Mathioudakis2010}. Yang and Leskovec found the cluster centroids of patterns associated with overall items and how the content's popularity grows and fades over time~\cite{Yang2011}. Roy \emph{et al.} proposed a novel transfer learning framework to grasp sudden popularity trend bursts with temporal knowledge from social streams~\cite{Roy2013}. Kong \emph{et al.} explored the problems of real-time prediction of bursting hashtags. Although these works exploited general comprehension on the variance mechanisms of popularity evolution, most of them are designed for general popularity or a group of items online~\cite{Kong2014}. These models have limited effectiveness on predicting the individual popularity in social media.

The other paradigm is modeling the individual popularity by temporal factors or variables with intrinsic time information~\cite{Wu2016, Shen2014, He2014, Zhao2015}, while researchers notice that temporal information is useful cues to predict individual popularity on social media in recent years~\cite{Wu2016, Li2015, Zhao2015}. Wu \emph{et al.} proposed to predict popularity in social media by unfolding its contextual dynamics and incorporated temporal context into the prediction~\cite{Wu2016}. Zhao \emph{et al.} combined human reaction time and post infectiousness to build the theory of self-exciting point processes~\cite{Zhao2015}. Shen \emph{et al.} proposed a reinforced Poisson process to model explicitly the dynamic popularity of individual items based on the arrival time of attention~\cite{Shen2014}. He \emph{et al.} used comments as a time-aware bipartite graph for estimating the popularity of item~\cite{He2014}. Although these models succeed in using temporal information, the essential multi-scale characteristic of popularity dynamics is lacking of investigation.

Therefore, how to incorporate multiple time-scales information into the study of popularity prediction in social media is still an open research issue. This motivated us to design the proposed framework of temporal modeling.

\begin{figure*}
  \centering\includegraphics[width=6.8in]{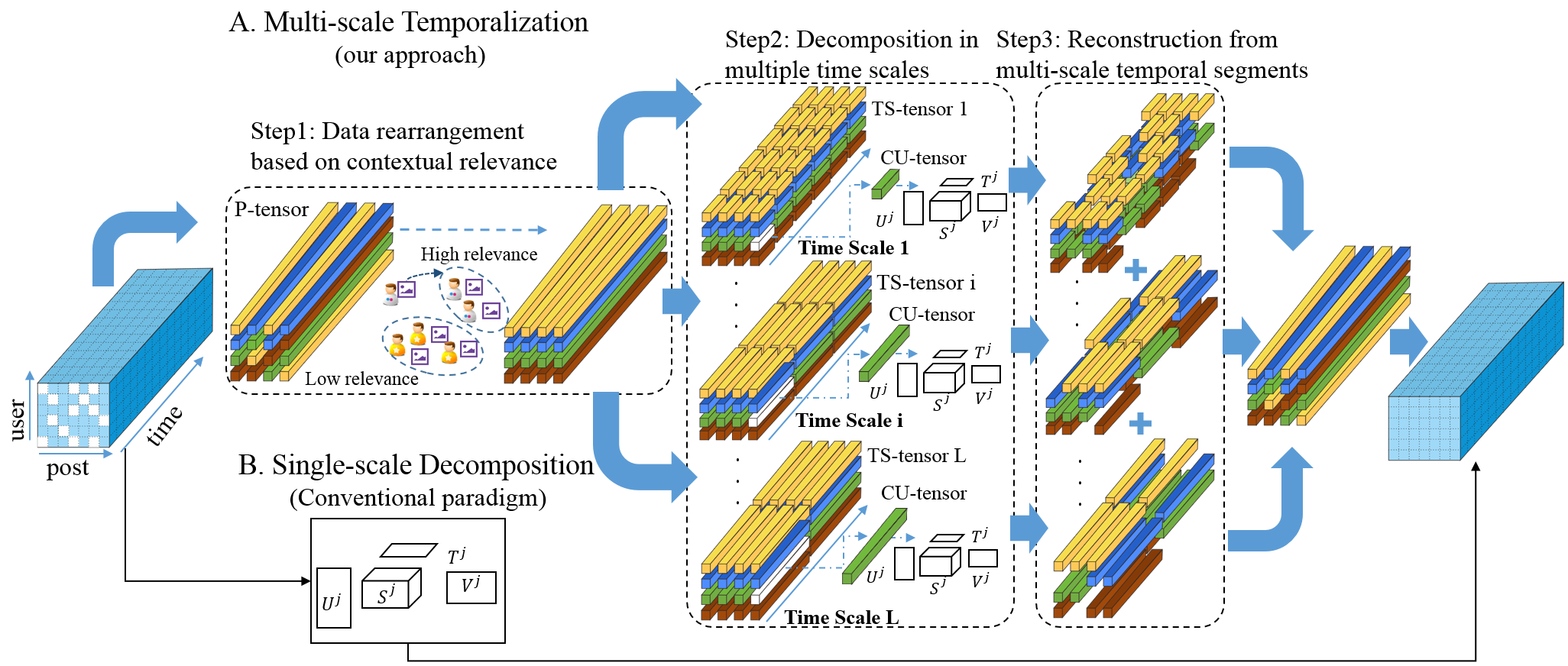}
  \caption{An overview of our proposed Multi-scale Temporalization (MT). In the first step, we proposed a data rearrangement step to re-organize data locations in tensor space with considering the corresponding contextual factors. Then we model temporal dynamics of popularity by multi-scale decomposition structures. Finally, we predict popularity with join low-rank constraints based on multi-level reconstruction.}\label{framework}
\end{figure*}

\section{Multi-scale Temporalization \\Modeling}
\subsection{Framework Overview}
In this section, we introduce the proposed multi-scale temporalization modeling for popularity dynamics. First, we formally define the problem of popularity prediction. Without loss of generality, we take photo posts on social media as our initial research focus.

\emph{Problem Formulation} \textbf{(Popularity Prediction)}: Given a group of users $U$ and a collection of photo posts $V$, the popularity of photos can be recorded as the collection $S$. Since the generation of popularity on social media is a collaborative process~\cite{Wu2016}, an individual popularity (the popularity of a specific post or item) $s$ corresponds to a sharing behavior that can be described by a three-dimensional tuple $\left\langle {u, v, t} \right\rangle $ denoting that a user $u$ shared a new photo $v$ at time $t$. A mathematical representation of the popularity data is defined as a three-dimensional data matrix, called popularity tensor $\mathbf{R}$, with the user dimension, photo dimension, and time dimension. The popularity prediction problem can thus be formulated as predicting unobserved popularity entries (with sharing behaviors that are unknown or not happened yet) in the tensor $R$ based on other observed data (entries with known popularities).

In Figure~\ref{framework}, we illustrate the proposed framework of multi-scale temporalization for popularity dynamics. Unlike traditional methods, our temporalization is a novel technique to unfold the temporal dynamics of popularity over time at multiple time scales, which includes three main analysis steps: (1) data rearrangement based on contextual relevance, (2) decomposition in multiple time scales, and (3) reconstruction from multi-scale temporal segments. Firstly, since the popularity data with similar contexts (e.g., users with similar profiles or photos with similar contents) tend to gain similar levels of popularity on social media~\cite{McParlane2014}, we group the popularity data with high contextual relevance to be in close proximity, cf. Section 3.2. Next, we perform multi-scale tensor decomposition to derive an approximate representation of the popularity data at different time scales. For each time scale (e.g., a scale of day, week, or month), the corresponding scale tensor is divided into a sequence of temporal segments with multiple time-scale representations, cf. Section 3.3. Finally, we perform a joint optimization strategy to combine a set of the temporal segments from all the time scales to best approximate the rearranged popularity tensor, where joint low-rank constraints are applied, cf. Section 3.4. As a consequence, estimates of the unobserved values in the popularity tensor can be effectively obtained.

\subsection{Data Rearrangement based on Contextual Relevance}
The popularity of a post is context-aware signal, which is highly correlated with certain post contexts, e.g., who shared the post, what the post is about, etc. As suggested by the previous findings~\cite{Cappallo2015, Gelli2015, Denton2015}, the posts with relevant contexts tend to gain similar popularity~\cite{McParlane2014, Jiang2014, Jin2016}. In the paper, our purpose is to partition a popularity tensor as a disjoint collection of smaller tensors, and predict new popularity in popularity tensor via pattern reconstruction on these tensors. However, uncovering patterns of the popularity data is difficult because the noise caused by context inconsistency (such as contextual differences in user factors or content factors) behind the temporal variation would result in analysis challenges. That is, if two neighboring data regions are contextually correlated, there would exist context-related noise in the popularity tensor prediction. Hence we hope popularity data in neighboring regions would have high correlation of the contextual information, such that each smaller tensor can help construct a ``compact'' representation for the corresponding popularity data in prediction.

Therefore, we propose a approach of contextual data rearrangement for the challenge in popularity temporalization. Most of previous works apply sorting-based methods to perform element-wise rearrangement in a matrix, but the high correlation of neighboring local data might not be guaranteed if multiple dimensions (e.g., user, photo, etc.) are involved in the sorting process. As inspired by~\cite{Gu2009}, we group the given tensor data in the multi-dimensional feature space without enforcing to sort the data. The proposed data rearrangement can help us to obtain the data units with multi-dimensional contextual consistency for more effective CU-tensor decomposition.

\begin{figure}
  \centering\includegraphics[width=3.3in]{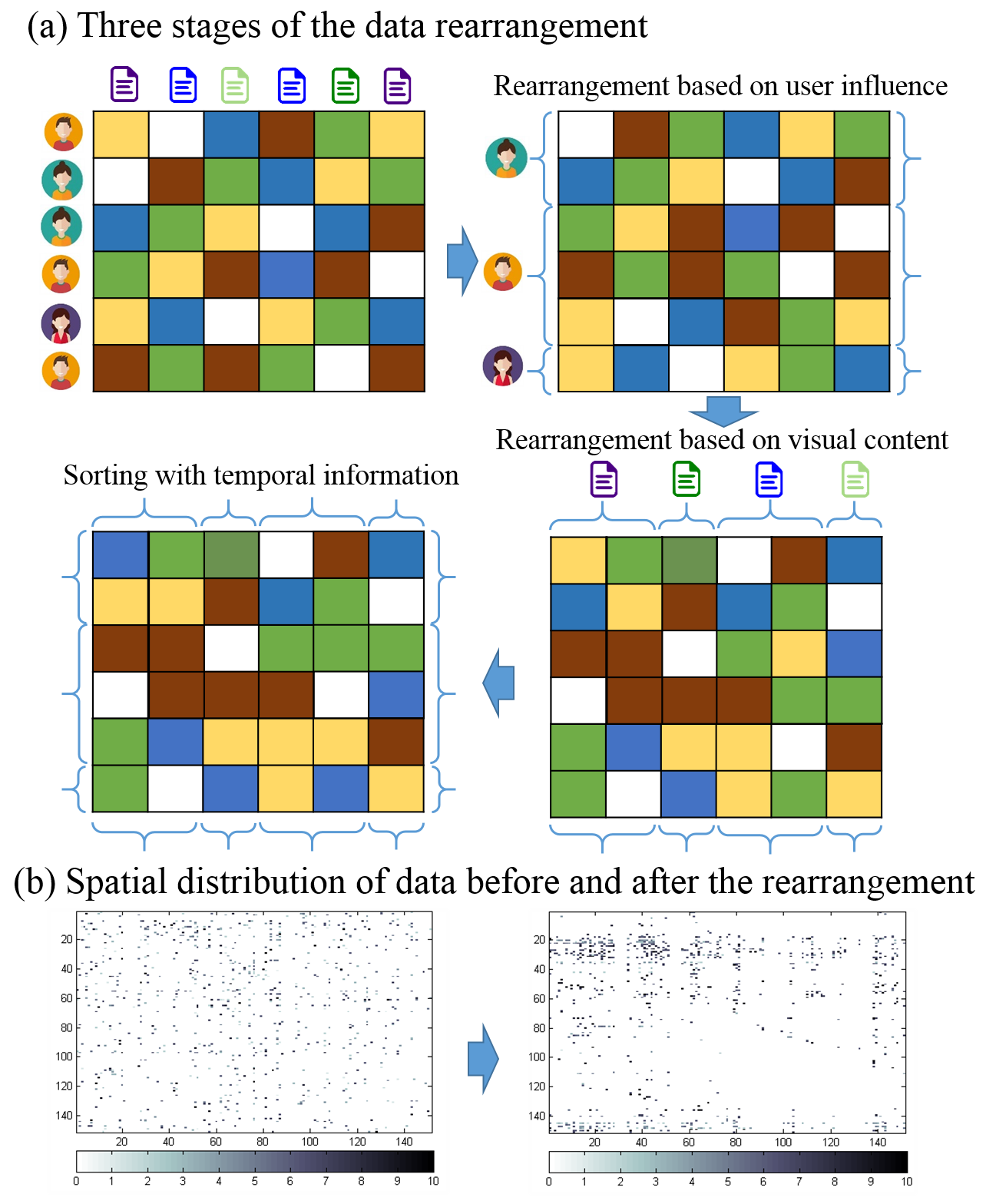}
  \vskip -10pt
  \caption{(a)The three main steps of the data rearrangement. (b) After the rearrangement is applied, neighboring data have higher contextual relevance.}\label{datarearrangement}
  \vskip -6pt
\end{figure}

To extract the contextual information of individual popularities, contextual features are built from three main perspectives: user influence, visual content, and post metadata.

\textbf{User Influence.} User influence reflects the personal reputation on social media, and the user features we adopted include: mean number of views, photo count, number of contacts, average number of members in a user's groups, and having a Pro Flickr account or not.

\textbf{Visual Content.} To describe photo content, we extract both low-level and object-level vision features. Low-level vision features are adopted to measure the color distribution and image local structures, including color patch descriptors~\cite{Khan2013}, Local Binary Pattern (LBP) descriptors, and locality-constrained gradient descriptors~\cite{Khosla2014, Dalal2005}. For object-level vision features, we use deep learning features by convolutional neural networks (CNNs) with the DeCAF method \cite{Donahue2013} to determine object categories in photos, resulting in 4,096 feature dimensions.

\textbf{Post Information.} Post information can be directly extracted from the metadata of photo files~\cite{Gelli2015}, containing the sharing time, the number of image tags, the length of the post title, and the length of post descriptions.

After the above context factors being extracted, the popularity tensor is rearranged in a three-step manner. Since the user influence has been shown to be a dominant factor in photo popularity prediction~\cite{Wu2016, Gelli2015, Denton2015}, the popularity data are first grouped along the user dimension by the k-means algorithm~\cite{Wagstaff2010} based on the user influence features. Note that the exact spatial order of either the groups or the data inside a group is not important because the purpose of our grouping here is to place ``similar'' data as close as possible but not to rank the data. As the second step, another grouping is performed along the post dimension based on the features of both the visual content and the post information (excluding the sharing time). Finally, for each data group (on the post dimension) of a specific user, we sort the data inside the group by the sharing time (from the post information) as a refinement step, cf. Figure~\ref{datarearrangement}.

\subsection{Decomposition in Multiple Time Scales}
We try to predict the popularity by the joint approximation with considering multiple time scales together. To capture the local dynamics of popularity over time, the proposed multiple time-scale decomposition operates by decomposing a popularity tensor (P-tensor) into a number of time-scale tensors (TS-tensors) each composed of a set of compositional unit tensors (CU-tensors), as detailed below. Conceptually, we attempt to break down the popularity variation into multiple levels of detail for constructing an accurate representation of popularity for popularity prediction.

Let $T=\{T_i\}_{i=1}^L$ denotes a set of adopted time scales. Given a popularity tensor $\mathbf{R}$ associated with a collection of users $U$ and a collection of photo posts $V$, the multiple time-scale decomposition can be expressed as the sum of $L$ components with each corresponding to an adopted time scale:
\begin{equation}
\mathbf{R} =\sum\limits_{i = 1}^L {{ \mathbf{R_i} }},
\end{equation}
where $\mathbf{R_i}$ is the same size of $\mathbf{R}$ and called a TS-tensor of $\mathbf{R}$ in the $T_i$ time scale. Since data with high contextual relevance are rearranged to be spatially close in $\mathbf{R}$ (cf. Section 3.2), each TS-tensor can be further partitioned into smaller pieces called CU-tensors as the unit of analysis, in order to take benefit of the data homogeneity for decomposition. That is, each CU-tensor in $\mathbf{R_i}$ has the dimensions of $m_j \times n_j \times t_j$, where the time dimension $t_j$ is a base-2 or base-$e$ number related to the time scale of $\mathbf{R_i}$ as discussed in the experiments. Then the user dimension $m_j$ and the post dimension $n_j$ are set to be the same order of magnitude as $t_j$. Note that $m_j$ (or $n_j$) needs to be larger than the smallest group size of users (or posts) as obtained in the data rearrangement.

The exact locations of CU-tensors in $\mathbf{R_i}$ can thus be indexed by a location set $M=\{M_j\}_{j=1}^{J_i}$ where $J_i$ is the total number of CU-tensors in $\mathbf{R_i}$. Thus $\mathbf{R_i}$ can be formulated as a summation of the CU-tensors by defining a reshaping transformation below.

Definition 1. \textbf{(CU-tensor Transformation)} Given an TS-tensor $\mathbf{R_i}$, the CU-tensor transformation is defined as an operation $O_i$ for converting one of $\mathbf{R_i}$'s CU-tensors, $\mathbf{R_i}^j$, $j = 1, ..., J_i$, into a zero tensor of the same size as $\mathbf{R_i}$ with only values of $\mathbf{R_i}^j$ in the corresponding location $M_j$, such that $\mathbf{R_i}=\sum\nolimits_{j=1}^{J_i}{O_i(M_j, \mathbf{R_i}^j)}$.

Therefore, the multiple time-scale decomposition of an original P-tensor can be simplified into a series of smaller-scale decompositions performed on the CU-tensors. In this paper, we applied the Singular Value Decomposition (SVD)~\cite{Chen2009,Gu2015} as our decomposition tool. The expression of $\mathbf{R}$ can be thus changed to:
\begin{equation}
\mathbf{R}=\sum\limits_{i=1}^L \sum\limits_{j=1}^{J_i}O_i(M_j, \mathbf{S_i}^j \times \mathbf{U_i}^j \times \mathbf{V_i}^j \times \mathbf{T_i}^j ),
\end{equation}
where $\mathbf{S_i}^j \times \mathbf{U_i}^j \times \mathbf{V_i}^j \times \mathbf{T_i}^j$ is the factorization of $\mathbf{R_i}^j$.

\subsection{Reconstruction from Multi-scale Temporal Segments}

In the previous section, we introduced the multi-scale decomposition process of temporalization for a P-tensor. The P-tensor can be further formed by a series of CU-tensor decompositions at multiple time scales. Since an individual popularity is related with a complex interplay among the contextual factors, where there often exists a high level of noise in the popularity data, making it a non-trivial task to conduct the popularity prediction~\cite{Roy2013, Shen2014}. Therefore, it is always a challenge about how to reduce the noise during recovering the P-tensor from multi-scale temporal segments.

For this challenge in the P-tensor reconstruction, we propose to apply joint low-rank constraints to reduce the noise level by minimizing the ranks of CU-tensors. More specifically, since the data rearrangement step (as proposed in Section 3.2) has aggregated the P-tensor entries with high contextual relevance in the data space, each CU-tensor has high contextual relevance among the data entries and the rank of the CU-tensor tends to be low. Therefore, we create joint low-rank constraints on CU-tensors instead of minimizing the gradients of the original objective function directly. Thus the optimization is formulated as:
\begin{equation}
\begin{array}{l}
\mathop {\min\quad }\limits_{{\mathbf{R_1} },...,{ \mathbf{R_L} }} \sum\limits_{i = 1}^L {\sum\limits_{j=1}^{J_i} {rank({ \mathbf{R_i} ^j})} } \\
{\rm{subject\; to\quad \mathbf{R} = }}\sum\limits_{i = 1}^L {{ \mathbf{R_i} }}.
\end{array}
\end{equation}
Especially, if the rank of a decomposed $\mathbf{R}^j_i$ is zero (calculated by SVD computation), we will ignore it for P-tensor reconstruction, which means the joint low-rank constraints will select the CU-tensors with a non-zero rank only in every optimization steps.

Since the minimization for the summation of rank constraints is a non-convex problem, we are unable to get an effective solution by optimizing it directly. Hence we convert the problem to a convex optimization problem by nuclear norm relaxation, which is inspired by the similar techniques applied in low-rank matrix completion~\cite{Koltchinskii2011}. Specifically, we construct the nuclear norm $|| \cdot |{|_{nuc}}$ to calculate the sum of singular values of overall elements of a CU-tensor $\mathbf{R_i^j}$, which is a minimization constraint of convex envelope of its rank. Meanwhile we have the nuclear norm $|| \cdot |{|^{(i)}}$ for each TS-tensor $\mathbf{R_i}$:
\begin{equation}
|| \cdot |{|^{(i)}} = \sum\limits_{j=1}^{J_i} {||{O_j}( \cdot )|{|_{nuc}}}.
\end{equation}

Then we incorporate the regularization term $||{ \mathbf{Z_i} }|{|^{(i)}}$ to obtain an equivalent relaxation formulation of our problem with two optimization objectives: a) the indicator constraint $\bbbone\{\cdot\}$ for recovering $\mathbf{R}$ based on multiple time-scale information components (CU-tensors), and b) the summation of regularization terms $\sum\nolimits_{i = 1}^L {{\lambda _i}||{ \mathbf{Z_i} }|{|^{(i){\rm{ }}}}}$ as a nuclear norm objective for optimization on joint rank constraints. The final objective function to estimate the partially known P-tensor $\mathbf{R}$ is formulated by:
\begin{equation}
\begin{array}{l}
\mathop {\min\quad }\limits_{{ \mathbf{R_i} },{ \mathbf{Z_{\rm{i}}} }} {\rm{ }} \bbbone \{ \mathbf{R} = \sum\limits_{i = 1}^L {{ \mathbf{R_i} }} \}  + \sum\limits_{i = 1}^L {{\lambda _i}||{ \mathbf{Z_i} }|{|^{(i){\rm{ }}}}} \\
{\rm{subject\; to\quad }}{{ \mathbf{R} }_i} = {\mathbf{Z_i}},
\end{array}
\end{equation}
where the regularization parameter $\lambda _i$ is to adjust the weight of proximal regularizations $||{ \mathbf{Z_i} }|{|^{(i)}}$. Note that, we compute parameters as $\lambda _i = \sqrt{m_i} + \sqrt{n_i}$ in our experiments inspired by similar parameter estimation of the previous work~\cite{Bandeira2014}.

 Since Alternating Direction Method of Multipliers (ADMM) is a powerful algorithm for convex optimization problems~\cite{ADMM2010}, we apply it on solving our optimization. The solution of the function is based on the joint optimization of two objectives and the iterative computation of $\mathbf{R_i}$ and $\mathbf{Z_i}$. At each iteration of the reconstruction of P-tensor $\mathbf{R}$, we need to estimate TS-tensors $\mathbf{R_i}$ at first with the following alternative updating steps:
\begin{equation}
\begin{array}{*{20}{l}}
 { \mathbf{R_i} } \leftarrow ({ \mathbf{Z_i} }  + \frac{1}{L}( \mathbf{R} - \sum\limits_{i = 1}^L { \mathbf{Z_i} } )),\\
\end{array}
\end{equation}
\begin{equation}
\begin{array}{*{20}{l}}
{ \mathbf{Z_i} } \leftarrow {SVD}_{\lambda _i} ({ \mathbf{R_i} }).\\
\end{array}
\end{equation}
 ${SVD}_{{\lambda _i} }(\cdot)$ is the abbreviation of SVD method with the mentioned regularization parameters $\lambda _i$ for updating $\mathbf{Z_i}$.  ${SVD}_{\lambda_i}$ is a proximal constraint function of the regularization summation term, we define ${SVD}_{\lambda_i}(R_i)=Umax(\Sigma-\lambda_i,0)T^ \mathrm{T}$.

 After updating all TS-tensors in each iteration, we identify them by estimating the individual correlation between the updated versions and the corresponding training data in P-tensor. By this step, we select the components with a positive correlation for the next iteration of P-tensor estimation. The algorithm of our method is the combination of alternative updating processes in each iteration, as summarized in Algorithm~\ref{alg}.   	

\begin{algorithm}[htb]
\caption{The Algorithm of Multi-scale Temporalization}
\label{alg}
\begin{algorithmic}[1]
\Require
estimated tensor: $\mathbf{R}$; maximal interaction: $n$;
\State Initial $\mathbf{R_0}$ and $\mathbf{Z_0}$ randomly on three-dimensional space with temporal information;
\Repeat
\Repeat
\State Optimization update steps:

\begin{center}
{\footnotesize
$
\begin{array}{*{20}{l}}
{ \mathbf{R_i} } \leftarrow ({ \mathbf{Z_i} }  + \frac{1}{L}(  \mathbf{R} - \sum\limits_{i = 1}^L { \mathbf{Z_i} } ))\\
{ \mathbf{Z_i} } \leftarrow {SVD}_{{\lambda _i} }({ \mathbf{R_i} })\\
\end{array}
$
}
\end{center}

\State Update $\mathbf{R_i}$ and $\mathbf{Z_i}$ followed by Alternating Direction Method of Multipliers

\Until{each TS-tensor $\mathbf{R_i}$ has been estimated}

\State Select $\mathbf{R_i}$ by correlation computation

\State Compute results of objective function:

\begin{center}
{\footnotesize
$
\begin{array}{l}
\mathbf{R}=\sum\limits_{i=1}^L \sum\limits_{j=1}^{J_i}O_i(M_j, \mathbf{S_i}^j \times \mathbf{U_i}^j \times \mathbf{V_i}^j \times \mathbf{T_i}^j )
\end{array}
$
}
\end{center}

\Until{procedure convergence or the number of interactions is over $n$}
\end{algorithmic}
\end{algorithm}

\subsection{Popularity Formulation}
The popularity of a shared post is related to the interaction behaviors of user preference in social media. For example, tweets popularity can be represented by ``retweeted count'' on Twitter, and video popularity is ``visiting count'' on YouTube. On Flickr, when browsing a personal photo stream or image search results, users can view details of a photo content with its metadata through clicking photo thumbnails. In our prediction, we use ``viewing count'' to describe the photo popularity, since it is a significant indicator of public preference of a new photo.

To alleviate the large variation (i.e. the number of views of different photos varies largely from zero to millions), we apply the log-normalization approach~\cite{Khosla2014} on the popularity formulation. As a result, the log-normalized popularity of a photo can be defined as
\begin{equation}
\label{EqnPop} s = {\log _2}\frac{r}{d} + 1,
\end{equation}
where $r$ is the original view count of each published photo, and $d$ is the number of days since the photo was shared.

\section{Experiments}
In this section, we report three groups of experiments: (1) we compare our proposed method and state-of-the-art algorithms for popularity prediction on over 1.8 million photos from Flickr, and our method is shown to be more effective on popularity prediction, (2) we perform further experiments to validate our algorithmic choices, e.g., the proposed time scales are effective to decompose popularity dynamics of social media photos, (3) we demonstrate the temporal prediction ability of our approach for popularity prediction and showcase some sample results.

\subsection{Datasets}
 For validating our proposed approach on real-world data, we collected over 1.8 million photos from over 70K users in total for two large-scale photo collections: the photo-mix dataset and the user-album dataset. These settings represent two different scenarios in social media platform.

\textbf{Photo-mix Dataset (PmD)}: We use the Visual Sentiment Ontology (VSO) dataset~\cite{Borth2013} consisting of approximately 1.2M images from over 70K users. Each user has five published photos at least. This setting often occurs on image retrieval or photo streams where people can browse various kinds of photos. We put all the images of these users together and performed popularity prediction on the full data.

\textbf{User-album Dataset (UaD)}: We collected 600K photos from the personal albums of 400 different users, and organized the photos into training data and testing data randomly. This setting is built for the scenario of finding interesting photographers by published albums. The prediction for the dataset tends to use more photos for each user than PmD.

For providing results on both of ``big data'' and ``small data'', we selected photos to get smaller subsets from PmD and UaD in number of photos, i.e., 400K from VSO\_CC dataset (a set of Flickr images with Creative Common (CC) licenses of VSO), 800K from VSO\_full (a set of images associated with the full VSO) and 300K from UaD (this set include half images of UaD), respectively.

\subsection{Evaluation Metric}
To evaluate performances of comparisons, we use the median Spearman Ranking Correlation with 10 fold cross-validation in our following experiments. The correlation is used to identify and test the strength of a relationship between a predicted popularity set $\hat{P}$ and the actual popularity set $P$. Supposing the size of test sample set is $k$, the calculation of spearman ranking correlation $r_s$ can be expressed as:
\begin{equation}
 r_s = \frac{1}{k-1} \sum ^k _{i=1} \left( \frac{P_i - \bar{P}}{\sigma_P} \right) \left( \frac{{\hat{P}}_i - \bar{\hat{P}}}{\sigma_{\hat{P}}} \right),
 \end{equation}
where $\bar{P}$ and $\sigma_P$ are the mean and the variance of the corresponding popularity set.

\subsection{Baselines}
\textbf{Baseline for Comparing Methods:}
In order to compare with state-of-the-art models, we implemented the following approaches which can be applied into popularity prediction problem as different baselines.

\emph{Baseline 1:} \textbf{Average Views (AV).} Since similar photos tend to obtain similar popularity, the popularity of testing photos can be estimated by the average views of its top five similar photos from training data. The popularity $s_j$ can be formulated by $\frac{1}{k}\sum\limits_{n = 1}^k {{s_n}}$,where $k$ is the number of similar photos, $s_n$ is the popularity of each similar photo with $v_j$. If the number of similar photos is over five, we rank them by the posting time and select the most recent photos.

\emph{Baseline 2:} \textbf{Logistic Regression (LR)~\cite{Szabo2008}.} From the early research, Szabo \emph{et al.} used Linear Regression (LR) in predicting popularity of online content, and it is a important work on popularity prediction tasks as baseline model. The multiple factors are organized as feature vectors in using this model.

\emph{Baseline 3:} \textbf{Support Vector Regression (SVR)~\cite{Khosla2014}.} Khosla \emph{et al.} used Support Vector Regression (SVR) in prediction and incorporated user cues and photo content as feature vectors. A linear kernel is used in the SVR.

\emph{Baseline 4:} \textbf{Bipartite Graph (BG)~\cite{He2014}.} Bipartite Graph model is widely used in popularity prediction and ranking. Suppose $G=(<U\cup V>, E)$ is a bipartite graph, where the set $U$ and set $V$ represent users and items respectively, and edges $E$ are posting behaviors. We use a regularization term $R(f)$ as
\begin{equation}
  R(f) = \frac{1}{2}\sum\limits_{j = 1}^n {\sum\limits_{i = 1}^m {{\omega _{ij}}\left( {\frac{{f({u_i})}}{{\sqrt {d_i^u} }} - \frac{{f({v_j})}}{{\sqrt {d_j^v} }}} \right)} },
\end{equation}
where $w_{ij}$ is defined by the posting behaviors between users and items. $d_i^u$ and $d_j^v$ are the weighted degrees of photo $v_j$ and user $u_i$ for normalization, respectively.

\emph{Baseline 5:} \textbf{Temporal Matrix Factorization (TMF)~\cite{Kong2011}.} The decomposition of a whole tensor on large data is not practical because large data is often too sparse to find an optimal solution. Hence we adopted a temporal modeling approach to incorporate temporal extension~\cite{Koren2009} into a practical Non-negative Matrix Factorization~\cite{Kong2011} for prediction as one baseline. We combine the user-item interaction (${\mathbf{B}_{u}(t)}$), temporal terms (${\mathbf{B}_{i}(t)}$) and regular terms in the optimization objective:
\begin{equation}
\arg \mathop {\min }\limits_{{\mathbf{U,V}}} \left\| {{\mathbf{R}} - {{{\mathbf{B}}_u}(t)} - {{{\mathbf{B}}_i}(t)} -{\mathbf{UV}}} \right\|_F^2 + {\lambda _{U}}{\left\| {\mathbf{U}} \right\|^2} + {\lambda _{\mathbf{V}}}{\left\| {\mathbf{V}} \right\|^2}.
\end{equation}

Note that, we use the intrinsic time series as the temporal feature in the feature vectors for LR, SVR, BG and TMF.


\textbf{Baseline for the Effectiveness Evaluation of Time Scales:}
 Currently, we adopt three time scales in our approach, including a week of year, a month of year, and a season of year. In order to validate the use of the adopted time scales and get more insights into our approach on the prediction effectiveness while different time scales are applied, we add some specific time scales in the experiments, including 2, 3, and 5 days, and 4, 5, and 6 months. For abbreviation, the three adopted time scales are altogether called GT (general time scales). The specific day and month scales are called DST (day specific time scales) and MST (month specific time scales), respectively. Three baselines are then created below and evaluated on 100K UaD.

\emph{Baseline 6:} \textbf{GT-GT$_i$.} We evaluate the individual contribution of each time scale in GT by removing a time scale from GT one at a time.

\emph{Baseline 7:} \textbf{GT-DST$_i$.} We extend the GT by adding a time scale from DST one at a time.

\emph{Baseline 8:} \textbf{GT-MST$_i$.} We extend the GT by adding a time scale from MST one at a time.

 \subsection{The Comparison of Prediction Methods}

{
\begin{table}[t]
\centering
  \caption{Prediction performances on PmD and UaD datasets (metric: Spearman Ranking Correlation).}\tabcolsep4pt \label{exp_method}
\footnotesize
\begin{tabular}{m{58pt}<{\centering} m{33pt}<{\centering} m{33pt}<{\centering} m{33pt}<{\centering} m{33pt}<{\centering}}\hline
\multicolumn{1}{c}
{\textbf{Method}}&\textbf{400K PmD}&\textbf{800K PmD}&{\textbf{300K UaD}}&\textbf{600K UaD} \\ \hline
AV                              &  0.0419           & 0.0785            &  0.1043           & 0.2022    \\
LR~\cite{Szabo2008}             &  0.2805           & 0.3026            &  0.5582           & 0.6359   \\
SVR~\cite{Khosla2014}           &  0.2627           & 0.2725            &  0.4415           & 0.6273    \\
BG~\cite{He2014}                &  0.2385           & 0.2573            &  0.2851           & 0.5465    \\
TMF~\cite{Kong2011}             &  0.2186           & 0.2839            &  0.4352           & 0.5795    \\
MT (ours)                       & \textbf{0.2916}   & \textbf{0.3264}   &  \textbf{0.5972}  & \textbf{0.6574}\\
\hline
\end{tabular}
\end{table}
}

 In Table~\ref{exp_method}, we give the prediction performance of our approach and the comparing methods. In each column of the table, our approach achieves the best performances. Especially in 800K PmD and 300K UaD, the relative improvements of our approach over the best baseline model LR is about 7.9\% and 6.9\%. Unlike TMF, which is also a temporal model based on matrix decomposition technique, our method considers multiple time-scale structure of popularity dynamics and even achieves a better relative increase about 13.4\%--37.2\%. That suggests the decomposition with multiple time scales is more adequate for utilizing temporal information. Besides, the models of LR and SVR generally outperform the temporal model TMF. It might imply if the temporal information is not well utilized, the resultant performance can be even worse than the scenario when temporal information is excluded. Meanwhile, Average Views is the worst and its best correlation is only about 0.2, it ignores abundant contextual data of popularity and results in limited prediction.

From the comparison with the different methods, the correlation of experiments on large dataset of PmD or UaD tends to obtain a better performance. This is consistent with the intuition that the prediction performance depends heavily on the numbers of data, and ``big data'' provides better prediction ability. Moreover, Table~\ref{exp_method} shows that the performances of prediction on UaD can be as high as up to 0.6574, which are higher than on PmD. This indicates that the prediction on less photos per user (photo-mix application) is more difficult even on a large dataset (the most performances are about 0.3 even on 800K photos).

Overall our approach achieves the best performance and gives more relative rank correlation than the other four comparing methods (excluding AV) in a relative improvement of 20.3\%--109.5\% (with BG), 15.0\%--37.2\% (with TMF), 4.8\%--35.3\% (with SVR) and 3.4\%--7.9\% (with LR). The average improvement of our approach over the four methods is thus from 10.9\%--47.5\%. It suggests that the temporalization technique are more effective on the prediction task.

\subsection{Evaluation of Time Scales for Popularity Prediction}

The results of using our predefined time-scales are shown in the right column of Table~\ref{single_scale}, where the highest ranking correlation is 0.5342 when using all GT in multi-scale decomposition. It indicates that using suggested general time scales is effective in unfolding temporal dynamics of popularity on social media. Furthermore, the other ``leave-one-out'' results are to validate the importance of each time scale. The result with the absence of season time scale achieves the lowest ranking correlation. That means the season-level time scale is more crucial than the others for photo popularity prediction. From the observation on the overall result, all the leave-one-out results are lower than using all the time scales in GT, implying that all the currently adopted time scales are important.

{
\begin{table}[t]
\centering
  \caption{Prediction Performances of predicting popularity based on GT without GT$_i$ (metric: Spearman Ranking Correlation).}\tabcolsep0.05pt \label{single_scale}
\footnotesize
\begin{tabular}{m{38pt}<{\centering} | m{55pt}<{\centering}m{58pt}<{\centering} m{59pt}<{\centering} m{30pt}<{\centering} }\hline

{\textbf{Settings}}&{\textbf{GT-GT$_{week}$}}&\textbf{GT-GT$_{month}$}&\textbf{GT-GT$_{season}$}&\textbf{GT}  \\ \hline
\textbf{$r_s$} &  0.3176  &  0.3578 &  0.1759  &  \textbf{0.5342}  \\

\hline
\end{tabular}
\end{table}
}

Next, we analyze the improvements of performance in terms of DST and MST, respectively. Compared with the performance of GT in Table~\ref{single_scale}, the performances in Table~\ref{unusual_scale} drop down by 1.3\%--8.5\%. Also, we found that MST prediction using the same data performs more accurately than using the DST (Table~\ref{unusual_scale}). This finding illustrates that the monthly temporalization of photo popularity dynamics on Flickr are more effective than day based analysis.

 {
\begin{table}[t]
\centering
  \caption{Prediction performances of GT by incorporating DST or MST (metric: Spearman Ranking Correlation).}\tabcolsep4pt \label{unusual_scale}
\footnotesize
\begin{tabular}{m{30pt}<{\centering} m{55pt}<{\centering}| m{45pt}<{\centering} m{60pt}<{\centering} }\hline
\multicolumn{1}{c}
{\textbf{GT+DST$_i$}}&{\textbf{Performance}}&\textbf{GT+MST$_i$}&\textbf{Performance} \\ \hline
2(D)  &  0.4485  & 4(M)   &  \textbf{0.5208} \\
3(D)  &  \textbf{0.4845}  & 5(M)   &  0.5142 \\
5(D)  &  0.4688  & 6(M)   &  0.4553 \\

\hline
\end{tabular}
\end{table}
}

\begin{figure*}
  \centering\includegraphics[width=6in]{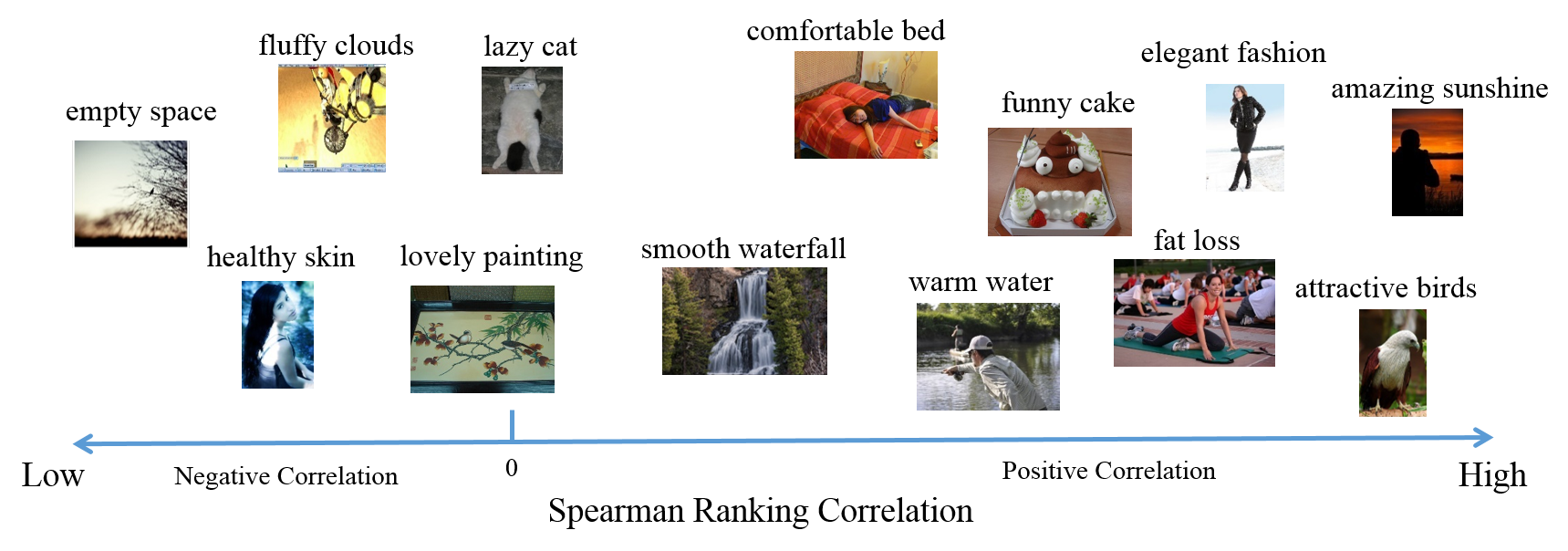}
  \vskip -10pt
  \caption{Sample photos with annotated image tags. In comparison to state-of-the-art methods, our approach takes into account both multi-scale temporal and contextual information. Accordingly, we can obtain higher correlation not only on the photos with clear temporal characteristic (e.g. ``amazing sunshine'' photos) but also on other common photos (e.g. ``attractive birds'' photos).}\label{correlation}
  \vskip -6pt
\end{figure*}

\subsection{Sample Results}
The purpose of this experiment is to observe the time-sensitive characteristic of our model by predicting the average ranking correlation of different photo categories. Sample photos with the associated image category tags are illustrated in Figure~\ref{correlation}. It can be found that our model does perform better for photos with time-related tags (e.g., ``amazing sunshine'') than those with general tags (e.g., ``healthy skin''). According to the results, our model have good prediction performance on time-related photos, such as those tagged with ``amazing sunshine'', ``fat loss'' or ``elegant fashion'' (the performance on several photos is even as high as up to over than 0.6 in prediction). The content of these photos are related with temporal natural evolution, user activity or public preference~\cite{Gan2015}. Besides, the popularity of those with general category tags in the left part of Figure~\ref{correlation} are difficult to predict, such as ``empty space'', ``fluffy clouds'' etc. We found that three observations may lead to the results. Firstly, the popularity of photos are varying without obvious patterns even on content, such as ``empty space'' and ``healthy skin'' in Figure~\ref{correlation}. That means the image tags are limited representations on abstract categories of photos. The second one is that some photos are blurred. This is also an unavoidable challenge not only to our model but also to others. Furthermore, several categories of photos like ``lazy cat'' or ``lovely painting'' have no obvious temporal patterns or even without explicit semantics for comprehension. This is a limitation of our model but might be improved in future works by incorporating into the text description content.

\section{Conclusions and Future Work}
In this paper, we have presented a general framework named Multi-scale Temporalization for modeling dynamic popularity in social media from multiple time scale view. Specially, we propose a structured decomposition model at different time-scales instead of computation on whole popularity tensor directly with only intrinsic time information. The experimental results showed that our approach outperform the state-of-the-art methods, with a relative improvement of averagely 10.9\%--47.5\% of our approach, which demonstrates the effectiveness on predicting power.

There are several possible directions for future investigation on temporal popularity prediction. One is to exploit new clustering strategies for capturing contextual relevance in data rearrangement with temporal factors. Another open question would be discovering prospective strategies for popularity in evolving systems. Furthermore, to unfold evolving popularity online content for detecting or predicting the user activity or influence is also essential for exploring social signal in social media.


\section{Acknowledgements}
This work is supported by National High Technology Research and Development Program of China (2014AA015202), National Nature Science Foundation of China (61525206, 61571424, 61428207) and Beijing Advanced Innovation Center for Imaging Technology (BAICIT-2016009).


%

\balance
\bibliographystyle{new}

\bibliography{PopularityPredictionMM16}  
%
%


\end{document}